\documentclass[11pt,a4paper]{article}
\usepackage{jheppub}
\usepackage{amsmath}
\usepackage[most]{tcolorbox}
\usepackage{dsfont}
\usepackage{ulem}
\usepackage{natbib}
\usepackage{xcolor}
\usepackage[hang,flushmargin]{footmisc}
\usepackage{tikz-cd}
\usepackage{enumitem}

\makeatletter\renewcommand{\@biblabel}[1]{#1.}\makeatother

\def\be{\begin{equation}}
\def\ee{\end{equation}}
\def\i{{\rm i}}

\def\e{{\rm e}}
\def\d{{\rm d}}

\def\C{{\mathbb{C}}}

\newcommand{\mb}[1]{\mathbf{#1}}

\def\ket#1{{ |#1\rangle}}

\def\res#1{\mathop \text{Res}\limits_{#1}}

\preprint{
{\small{\textsf{}}}}

\title{Bootstrapping the $S^5$ partition function}

\author[a]{Fabrizio Nieri}
\author[b]{Yiwen Pan}
\author[a]{Maxim Zabzine}

\affiliation[a]{Department of Physics and Astronomy, Uppsala University,\\
Box 516, SE-75120 Uppsala, Sweden.}
\affiliation[b]{School of Physics, Sun Yat-Sen University,\\
Guangzhou, Guangdong, China.}

\emailAdd{fb.nieri@gmail.com}
\emailAdd{panyw5@mail.sysu.edu.cn}
\emailAdd{maxim.zabzine@physics.uu.se}

\abstract{We consider  $U(N)$ SQCD on $S^5$ and propose a Higgs branch-like expression for its partition function. We support the result by arguing that the knowledge of certain BPS codimension 2 and 4 defects arising from Higgsing is enough to reconstruct the bulk partition function, and that the defect partition functions satisfy a set of non-perturbative Schwinger-Dyson equations. We show that the result is consistent with, and naturally come from, the BPS/CFT perspective. In this language, the defect partition functions are identified with free boson correlators of the $q$-Virasoro modular triple, and the constraint equations with Ward identities satisfied by the corresponding Dotsenko-Fateev $q$-conformal blocks, providing a natural basis to expand the  $S^5$ partition function.
}


\begin{document}
\maketitle

\flushbottom

\section{Introduction}
In recent years, localization has been an amazing tool for non-perturbative studies of SUSY gauge theories in various dimensions and backgrounds (see \cite{Teschner:2016yzf,Pestun:2016zxk} for a review). Compact manifolds  are ideal spacetimes because they provide natural infrared regulators and allow more straightforward application of the relevant mathematical techniques.

The localized BPS observables of SUSY gauge theories on compact spaces, such as the partition functions, are usually presented as matrix-like integrals over the Cartan subalgebra of the gauge group, a scheme a.k.a. {\it Coulomb branch localization}. However, scheme independence implies that these observables may have completely different but equivalent expressions, and one presentation can be better than (or complementary to) another depending on the purpose. In fact, another popular localization scheme is {\it Higgs branch localization}, in which case localized BPS observables are typically presented as summations over discrete sets of solutions to certain vacuum equations. In lower dimensions, it is often possible to explicitly show the equivalence of the two approaches by brute force residue calculus (see e.g. \cite{Pasquetti:2011fj,Doroud:2012xw,Benini:2012ui,Fujitsuka:2013fga,Benini:2013yva,Taki:2013opa,Peelaers:2014ima,Yoshida:2014qwa}). However, this is not the case in higher dimensions, where the presence of instantons makes life much harder. The study of the alternative Higgs branch localization scheme in higher dimensions has begun only recently \cite{Pan:2014bwa,Pan:2015hza,Chen:2015fta}. Remarkably, this approach has provided a new window into a class of codimension 2 and 4 defects of 4d $\mathcal{N}=2$ gauge theories on the 4-sphere $S^4$ \cite{Pestun:2007rz,Hama:2012bg}, including their AGT duals \cite{Alday:2009aq,Wyllard:2009hg} in Liouville/Toda 2d CFTs \cite{Gomis:2016ljm}. 

In this note, we focus on SUSY gauge theories on the 5-sphere $S^5$ \cite{Kallen:2012va,Kallen:2012cs,Kim:2012qf,Hosomichi:2012ek,Imamura:2012bm,Lockhart:2012vp}. The main goal is to provide a  Higgs branch-like expression for the partition function of the 5d $\mathcal{N}=1$ $U(N)$ SQCD, closely following \cite{Pan:2016fbl}. This concrete choice is due to the simple brane realization of the theory \cite{Aharony:1997bh,Aharony:1997ju} and the known BPS/CFT and 5d AGT description \cite{Awata:2009ur,Awata:2010yy,Awata:2011dc,Mironov:2011dk,Nekrasov:2012xe,Nekrasov:2013xda,Carlsson:2013jka,Nieri:2013yra,Nieri:2013vba,Aganagic:2013tta,Aganagic:2014kja,Aganagic:2014oia,Zenkevich:2014lca,Mitev:2014isa,Isachenkov:2014eya,Aganagic:2015cta,Kimura:2015rgi,Benvenuti:2016dcs,Bourgine:2016vsq,Bourgine:2017jsi,Aganagic:2017smx} (before and after fiber/base duality \cite{Katz:1997eq,Bao:2011rc}). However, most of what we discuss for this model is expected to generalize to quiver theories. One of our guiding principle is that the it should be possible to reconstruct the $S^5$ partition function from the knowledge of certain BPS defects, very much in the spirit of \cite{Gaiotto:2012xa} for the 4d superconformal index  (see also \cite{Chen:2014rca}), and that the defect partition functions should have a dual 2d CFT-like interpretation (see e.g. \cite{Alday:2009fs,Bonelli:2011wx,Bonelli:2011fq,Gomis:2014eya,Bullimore:2016hdc} in four dimensions). While our Higgs branch result formally follows from the residue calculation of the matrix integral, we do not have a rigorous proof of its equivalence with the Coulomb branch expression due to some assumption on the contributing poles. However, we support our proposal by showing that it has all the expected properties, both from the the gauge theory and 5d AGT perspectives: $i$) the partition function has poles at special values of the mass parameters due to the pinching mechanism of the integration contour, and taking the residues can be interpreted as RG flows to defect theories; our proposal indeed assumes the form of a discrete sum whose summands capture the partition functions of codimension 2 and 4 defects which are consistent with the Higgs phase from the brane picture; $ii$) our proposal looks like the compact space version of a similar result in the 5d $\Omega$-background \cite{Nieri:2017ntx}, which can be proven combinatorially; $iii$) the defect partition functions can manifestly be constructed through the {\it $q$-Virasoro modular triple} \cite{Nieri:2017vrb}, in agreement with \cite{Nieri:2013yra,Nieri:2013vba}; quite interestingly, it turns out that the free boson correlators form a natural basis for solutions to $q$-deformed Ward identities or $q$-conformal blocks, and a similar approach has been recently used to study $\mathcal{W}_3$ conformal blocks of Toda 2d CFT \cite{Coman:2017qgv}.


The rest of this note is organized as follows. In section \ref{SUSYgauge}, we briefly review part of the geometry of the odd dimensional spheres we are interested in as well as the Coulomb branch matrix model of the $U(N)$ SQCD on $S^5$, and we give our proposal for its Higgs brach representation. In section \ref{ALGEBRAside}, we summarize the $q$-Virasoro modular triple construction and show its relevance to the physics on $S^5$. In section \ref{sec:disc}, we comment on open questions and outlook. Due to space limitations, we could not provide all the relevant background to both subjects, most of which can be found in the cited literature. For the same reason, we could not include technical appendices summarizing standard gauge theory results and definitions of various special functions, which can be found in the references.

\section{The supersymmetric gauge theory}\label{SUSYgauge}

\subsection{The geometry}

\begin{figure*}
  \centering
  \includegraphics[width=0.7\textwidth]{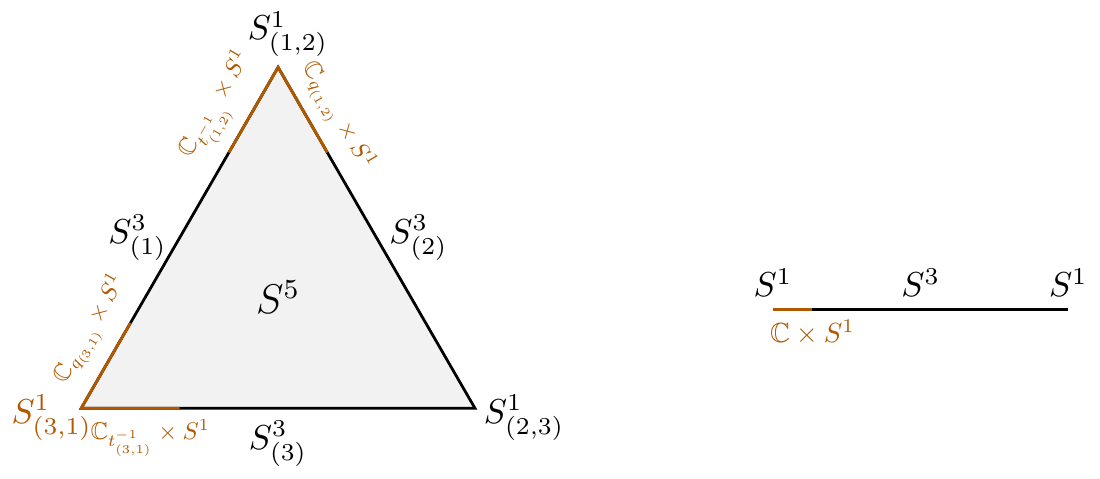}
  \caption{\label{Torusfibration} The squashed $S^5$ (resp. $S^3$) is a $T^3$ (resp. $T^2$) fibration over a solid triangle (resp. an interval). We also indicate several tubular neighborhoods of the $S^1_{(i,j)}$ inside the three $S^3_{(i)}$, whose $\Omega$-deformation are determined by the equivariant parameters.}
\end{figure*}

 The round $S^5$ defined by equation $|z_1|^2+ |z_2|^2 + |z_3|^2 = 1$ in $\C^3$ is a toric Sasaki-Einstein manifold equipped with a canonical contact structure and an associated Reeb vector. A refinement is provided by the squashed $S^5$, which is defined through a deformation of the Reeb vector specified by the real part of three complex squashing parameters $\omega_1, \omega_2, \omega_3$. Similarly, the squashed 3-sphere $S^3$ comes with a single squashing parameter $b$ (we refer to \cite{Qiu:2014oqa,Hama:2011ea} for further details on these geometries). Due to the toric nature, $S^5$ (resp. $S^3$) can be glued out of elementary patches $\mathbb{C}^2 \times S^1$ (resp. $\mathbb{C} \times S^1$) sitting at the corners of a torus $T^3$ (resp. $T^2$) fibration over a solid triangle (resp. interval). The three edges of the triangle correspond to three submanifolds $S_{(i)}^3 \subset S^5$ specified by the condition $z_i = 0$, where the 5d Reeb vector induces the 3d Reeb vectors on $S^3_{(i)}$ with squashing parameters $b_{(i)} \equiv \sqrt{\omega_{i+1}/\omega_{i+2}}$. The three $S^3_{(i)}$ intersect at three separate circles $S^1_{(i,j)} \equiv S^3_{(i)} \cap S^3_{(j)}$, corresponding to the three vertices of the triangle. This is illustrated in Figure \ref{Torusfibration}.
For later discussion, in Table \ref{parameter-table} we have organized the squashing parameters into exponentials.
\begin{table}[h]
\centering
\caption{The 5d squashing parameters can be reorganized into a set of exponentiated parameters. Each column can be associated to a vertex of the triangle in Figure \ref{Torusfibration}, as indicated by the subscripts $(i,j)$.}
\label{parameter-table} 
\begin{tabular}{c|c|c|c}
$(i,j)$ & (1,2) & (2,3) & (3,1) \\
\hline
$q_{(i,j)}$ & $\e^{2\pi \i \frac{\omega_3 + \omega_1}{\omega_3}}$ & $\e^{2\pi \i \frac{\omega_1 + \omega_2}{\omega_1}}$  & $\e^{2\pi \i \frac{\omega_2 + \omega_3}{\omega_2}}$ \\
$t_{(i,j)}$ & $\e^{ - 2\pi \i \frac{\omega_2}{\omega_3}}$ & $\e^{ - 2\pi \i \frac{\omega_3}{\omega_1}}$ & $\e^{ - 2\pi \i \frac{\omega_1}{\omega_2}}$
\end{tabular}  
\end{table}
From the naming conventions, one can infer that each pair of parameters $(q,t)$ is associated to a vertex in Figure \ref{Torusfibration}. The pattern $q_{(i,j)} \equiv \e^{2\pi \i (\omega_k + \omega_i)/\omega_k}$, $t_{(i,j)} \equiv \e^{- 2 \pi \i \omega_j/\omega_k}$, with $(i,j,k)$ any even permutation of $(1,2,3)$, will be used throughout this note.

\subsection{The partition function}
The partition functions of 5d $\mathcal{N} = 1$ SUSY gauge theories on the squashed $S^5$ can be computed via localization \cite{Kim:2012qf,Hosomichi:2012ek,Imamura:2012bm,Lockhart:2012vp}. The result for $U(N)$ SQCD reads
\begin{align}
  Z^{S^5}_\text{SQCD}(M, \tilde M, g_\text{YM}) \equiv \int \d^N\Sigma \, Z_\text{cl}^{S^5} Z^{S^5}_\text{1-loop} Z^{S^5}_\text{inst} \equiv \int \d^N  \Sigma \, Z^{S^5}_\text{SQCD}(\Sigma) ~,
  \label{S5-partition-function}
\end{align}
where the classical and the 1-loop contributions are given by
\begin{align}
  Z^{S^5}_\text{cl} \equiv \e^{- \frac{  8\pi^3   }{  \omega_1 \omega_2 \omega_3 g^2_\text{YM}  }\text{tr} \Sigma^2}\ , \quad Z^{S^5}_\text{1-loop} \equiv \frac{\prod_{A \ne B} S_3(\i\Sigma_A - \i\Sigma_B)}{\prod_{A=1}^N \prod_{I = 1}^{2N} S_3(\i \Sigma_A - \i M_I + \omega/2)}  \ ,
\end{align}
where $g_\text{YM}$ is the Yang-Mills coupling, $M_I$ are the masses of (anti-)fundamental hypers (depending on the sign) and we defined $\omega\equiv \omega_1+\omega_2+\omega_3$. The definition and properties of the triple Sine function can be found in \cite{Narukawa:2003}. It can be shown that these pieces  respect the toric structure of $S^5$, meaning that they are factorizable into three factors capturing the contributions from the tubular neighbourhood $\mathbb{C}^2 \times S^1$ of the three toric fixed-points \cite{Lockhart:2012vp,Nieri:2013vba,Qiu:2013aga}. Since it is believed that this is a property of the full integrand, the instanton part $Z^{S^5}_\text{inst}$ is usually presented as the product of three 5d $\Omega$-background instanton partition functions \cite{Lossev:1997xxx,Moore:1998et,Moore:1997dj,Losev:1997tp,Nekrasov:2002qd,Nekrasov:2003rj} with equivariant parameters entering as in Table \ref{parameter-table}.

\begin{figure}[t]
  \centering
  \includegraphics[width=0.3\textwidth]{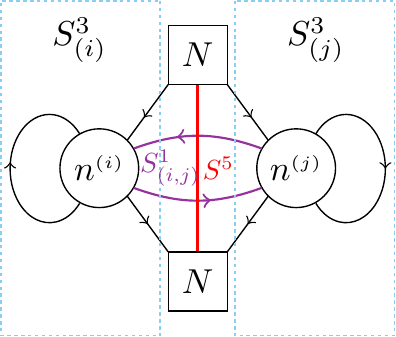}
  \caption{\label{intersecting-SQCDA} Here we depict part of the full quiver description of the intersecting gauge theory, where we include only the two 3d SQCDA's on $S^3_{(i)}$ and $S^3_{(j)}$ coupled to the 1d bi-fundamental chirals on the intersection $S^1_{(i,j)}$. For the full quiver one simply repeat this patter for all three pairs of 3d theories. The purple arrowed lines denote the 1d bi-fundamental chiral multiplets supported on the intersection $S^1_{(i,j)}$.}
\end{figure}

The integration contour in (\ref{S5-partition-function}) is along the real Cartan subalgebra of the gauge group parametrized by constant scalars $\Sigma_A$ in the vector multiplet. When all the external parameters are generic, one can close the contour by infinitely large half circles in the $\Sigma_A$ planes and try to explicitly evaluate the integral by collecting the residues of the enclosed poles. The real challenge is  to control them. We can essentially split the computation into two steps, namely the identification of the relevant class of poles and the summation over inequivalent classes. First of all, there are obvious poles in the 1-loop determinant at
\begin{align}\label{relpoles}
  \i\Sigma_A = \i\Sigma_A^* \equiv \i M_A  - \sum_{i=1}^3 n_A^{(i)}\omega_i - \frac{\omega}{2} \ , \qquad A = 1, \ldots, N \ ,
\end{align}
for masses $M_A$ of any $N$ out of the $2N$ fundamental hypers, and any set of non-negative integers $\{n^{(i)}_A\}$. We believe (and substantiate in the next section) that these are all the relevant poles that contribute to the final result.\footnote{It can be proven in the Abelian case, where instantons can be easily resummed.} We also let $M'_B$ denote the masses of the remaining $N$ hypers. A selection of $N$ out of the $2N$ hypers will be referred to as a \textit{Higgs vacuum}, written as HV for short. For this type of poles, it turns out that for each Higgs vacuum and three given non-negative integers $\{n^{(i)}\}$, the sum of the residues at the poles with fixed $\sum_{A = 1}^N n_A^{(i)} = n^{(i)}$ combine into something remarkably simple, namely
\begin{align}
  \sum_{\{\sum_A n_A^{(i)}=n^{(i)}\}} \res{\Sigma \to \Sigma^*} Z^{S^5}_\text{SQCD}(\Sigma) = Z^{S^5}_\text{cl} Z_\text{HM}^{S^5}\Big|_\text{HV} Z^{S^3_{(1)} \cup S^3_{(2)} \cup S^3_{(3)}}_{U(n^{(1)}), U(n^{(2)}), U(n^{(3)}),n_\text{f}=N}\Big|_\text{HV}\ ,
  \label{residue}
\end{align}
where $Z_\text{cl}^{S^5}|_\text{HV}$ is the classical exponential factor evaluated at the selected HV with $n^{(i)}_A=0$ and $Z_\text{HM}^{S^5}|_\text{HV}$ is the $S^5$ partition function of $N^2$ free hypers with masses $M_{AB}\equiv M_A - M'_B + \i\omega/2$
\be
Z_\text{HM}^{S^5}|_\text{HV}\equiv \prod_{A,B=1}^N\frac{1}{S_3(\i M_{AB}+\omega/2)}~.
\ee
The last factor in (\ref{residue}) can be interpreted as the partition function of a gauge theory on the {\it intersecting space} $S^3_{(1)} \cup S^3_{(2)} \cup S^3_{(3)}$, which deserves a bit of explanation. The gauge theory can be defined by placing the usual $U(n^{(i)})$ SQCDA\footnote{With the name $U(n)$ SQCDA, we refer to the $U(n)$ Yang-Mills theory coupled to (massive) $n_\text{f}$ fundamental, $n_\text{f}$ anti-fundamental and 1 adjoint chiral multiplets.} on each $S^3_{(i)}$ with the FI parameters turned on. The 3d degrees of freedom are further coupled to additional pairs of 1d chiral multiplets living on the intersections $S^1_{(i,j)}$ and transforming in the bi-fundamental representation of the neighboring 3d gauge groups $U(n^{(i)})$ and $U(n^{(j)})$. Such a gauge theory is nothing but the worldvolume theory of codimension 2 defects intersecting at codimension 4 loci arising from the baryonic Higgsing procedure described in \cite{Gaiotto:2012xa,Gaiotto:2014ina}. The masses of the adjoint chirals are determined purely by the geometry as $m_\text{adj}^{(i)} = \i \omega_i/\sqrt{\omega_j \omega_k}$, the $N$ fundamental masses $m^{(i)}_A$ (resp. anti-fundamental $\tilde m^{(i)}_B$) are determined in terms of $M_A$ (resp. $M'_B$)
\be
m_A^{(i)}=\frac{M_A+\i\omega_i}{\sqrt{\omega_j\omega_k}}+\i\frac{b_{(i)}+b_{(i)}^{-1}}{2}~,\qquad 
\tilde m_B^{(i)}=\frac{M'_B}{\sqrt{\omega_j\omega_k}}+\i\frac{b_{(i)}+b_{(i)}^{-1}}{2}~,
\ee
and finally the FI's are related to the 5d coupling as $\zeta^{(i)}_\text{FI}=8\pi^2 g_\text{YM}^{-2}/\sqrt{\omega_j\omega_k}$. As a result, there are relations between the masses\footnote{\label{mass-relations}In particular, $b_{(i)}m^{(i)}_A - b^{-1}_{(i+1)}m^{(i+1)}_A = - \frac{\i}{2}(b_{(i)}^2 - b_{(i+1)}^{-2}) ~,\quad  b_{(i)}\tilde m^{(i)}_A - b^{-1}_{(i+1)}\tilde m^{(i+1)}_A = + \frac{\i}{2}(b_{(i)}^2 - b_{(i+1)}^{-2})$.} across the three $S^3_{(i)}$,
signalling the presence of additional superpotential couplings between the subsystems. Schematically, the partition function can be organized into the elegant matrix model 
\begin{multline}\label{intmm}
  Z^{S^3_{(1)} \cup S^3_{(2)} \cup S^3_{(3)}}_{U(n^{(1)}), U(n^{(2)}), U(n^{(3)}),n_\text{f}=N}\Big|_\text{HV} \equiv\\
  \equiv \int\prod_{i=1}^{3}\frac{\d^{n^{(i)}}\sigma^{(i)}}{n^{(i)}!} \, Z_{U(n^{(i)}),n_\text{f}=N}^{S^3_{(i)}}(\sigma^{(i)}) \!\!\!\! \prod_{(i,j)\in\{(1,2),(2,3),(3,1)\}} \!\!\! Z_\text{chiral}^{S^1_{(i,j)}}(\sigma^{(i)},\sigma^{(j)}) \ ,
\end{multline}
where each term in the first factor encodes the classical and 1-loop contributions from the individual $U(n^{(i)})$ SQCDA on  $S^3_{(i)}$, and each term in the last factor represents the contribution from the 1d chiral multiplets on the circle $S^1_{(i,j)}$. Their explicit expressions and the integration contour\footnote{The contour is specified by a Jeffrey-Kirwan prescription.} can be found in \cite{Pan:2016fbl}. We emphasise that the choice of Higgs vacuum HV affects all the factors in (\ref{residue}). In particular, it determines the masses  in (\ref{intmm}). The resulting theory can be described by a quiver gauge theory, part of which is illustrated in Figure \ref{intersecting-SQCDA}.
The detailed presentation of the technical derivation of the result is beyond the scope of this short note due to space limitations. In a nutshell, this essentially follows from a few combinatorial manipulations relying on the fact that: $i$) the residues at the poles (\ref{relpoles}) of the $S^5$ 1-loop determinant can be written in terms of $S^3$ and $S^1$ determinants (double and single Sine functions) and a 5d remnant; $ii)$ at the same points, the SQCD instanton partition functions collapse to SQCDA vortex partition functions. We refer to \cite{Nieri:2017ntx} for analogous computations in the 5d $\Omega$-background, \cite{Pan:2016fbl,Nieri:2018xxx} for the intersecting space $S^3_{(i)}\cup S^3_{(j)}$ (i.e. the $n^{(k)}=0$ sub-case) and  \cite{Pan:2015hza,Pan:2016fbl} for the similar four dimensional setup. 

Finally, under our working assumptions, the full $S^5$ partition function can be simply obtained by summing the above matrix model over all possible ranks $\{n^{(i)}\}$ and HV weighted by the classical and free hyper contributions from the bulk. The final result reads
\begin{align}\label{finalresult}
  Z^{S^5}_\text{SQCD} = \sum_\text{HV} Z^{S^5}_\text{cl} Z_\text{HM}^{S^5}\Big|_\text{HV} \sum_{\{n^{(i)}\geq 0\}} Z^{S^3_{(1)} \cup S^3_{(2)} \cup S^3_{(3)}}_{U(n^{(1)}), U(n^{(2)}), U(n^{(3)}),n_\text{f}=N}\Big|_\text{HV} \ .
\end{align}

In the next section, we look at this proposal as a consequence of a dual algebraic description of the partition function and a set of identities which each residue has to satisfy. From this perspective, our working assumption of the contributing poles (\ref{relpoles}) can be completely justified, and the tedious technical derivation of (\ref{finalresult}) sketched above can be completely forgotten in favour of a much simpler algebraic construction. In order to do that, we first need to introduce the {\it generating function}.

\subsection{The generating function}
In \cite{Kimura:2015rgi}, Kimura and Pestun have considered an extended version of the pure $U(N)$  partition function in the 5d $\Omega$-background, which includes a set of infinitely many {\it time variables} $\{\tau_n,n\in\mathbb{Z}_{>0}\}$ as external parameters (see also \cite{Kimura:2016ebq} for a short introduction). We refer to it as the generating function $Z^{\mathbb{C}^2\times S^1}_{U(N)}(\tau)$. This refined object satisfies infinitely many differential constraints\footnote{In the matrix model context, these are also known as Virasoro constraints \cite{Morozov:1994hh}.} (of arbitrarily high degrees) in the time variables of the form
\be\label{qvircons}
T_m \, Z^{\mathbb{C}^2\times S^1}_{U(N)}(\tau)=0~,
\ee  
where $\{T_m,m\in\mathbb{Z}_{>0}\}$ are certain operators  built out from the time variables and their derivatives. Notice that the partition function is simply recovered at $\tau_n=0$. This family of operators can be enlarged to include additional operators $\{ T_{-m},m\in\mathbb{Z}_{\geq 0}\}$ which do not annihilate the generating function. Altogether, they can be assembled into a formal series $T(z)\equiv \sum_m T_m z^{-m}$, providing an interesting gauge theory observable, the $qq$-character \cite{Nekrasov:2015wsu},\footnote{For the relation to codimension 4 defects we refer to \cite{Kim:2016qqs,Assel:2018rcw}.} whose v.e.v. is regular as a function of $z$. Once this observable is given, one may define the generating function as the unique solution to the constraints (\ref{qvircons}), possibly once additional conditions are imposed. In this formalism, the inclusion of fundamental matter with flavor fugacity $\mu$ can be simply accounted for by a constant background for the time variables  of the form $\tau_n\to \tau_n+\mu^n/n$, of course resulting in modified constraints. Importantly, the inclusion of matter allows the theory to be Higgsed. In practice, the Coulomb branch parameters are set to special points where the instanton series truncates to a  vortex partition function, and it turns out that the generating function of the defect theory also satisfies analogous constraints \cite{Nedelin:2016gwu}. This observation has led us to the study of codimension 2 (and 4) expansions of the instanton series \cite{Nieri:2017ntx}, finding that the full $\Omega$-background partition function can be naturally assembled from lower dimensional pieces.

Our goal is to apply this reasoning to the $S^5$ partition function and to compute (or bootstrap) it by simpler means. In the next section, we show how to build the simplest and most natural solutions to constraint equations for the compact space setup, corresponding exactly to (\ref{intmm}). The key idea is to study and exploit the quantum group nature of the constraints.

\section{The algebra}\label{ALGEBRAside}

The constraint equations (\ref{qvircons}) can be identified with the Ward identities of the $q$-Virasoro or $\mathcal{W}_{q,t^{-1}}(A_1)$ algebra \cite{Shiraishi:1995rp,1997q.alg.....8006F} and $T(z)$ with its generating current. This follows from Kimura-Pestun's identification of the generating function with the state made out of infinitely many $q$-Virasoro screening charges in a Fock module over a reference vacuum of a ($q$-deformed) free boson. The map between the time variables and the usual Heisenberg oscillators $\{\mb{a}_n,n\in\mathbb{Z}^\times\}$ is roughly given by $\tau_n\sim \mb{a}_{-n}$, $\partial/\partial\tau_n\sim \mb{a}_n$, $n>0$, with the Fock vacuum identified with a $\tau$-independent function. This construction can obviously be extended to include an arbitrary number of commuting copies of the algebra. However, when considering three copies with deformation parameters entering as in Table \ref{parameter-table}, a peculiar construction, the $q$-Virasoro modular triple, arises \cite{Nieri:2017vrb}. In this section, we show how this algebraic setting naturally leads to  (\ref{finalresult}).

\subsection{The $q$-Virasoro algebra}

The $q$-Virasoro algebra can be defined as the associative algebra generated by the generators $\{\mb{T}_m, m\in\mathbb{Z}\}$ satisfying certain commutation relations (see e.g. \cite{Shiraishi:1995rp}). 
The algebra admits a free boson realization in terms of Heisenberg oscillators $\{\mb{a}_m,m\in\mathbb{Z}^\times\}$ and zero modes $\{\mb{P},\mb{Q}\}$ satisfying the non-trivial commutation relations 
\begin{align}
  & [\mb{a}_m, \mb{a}_n] = \frac{1}{n} (q^{n/2} - q^{-n/2})(t^{n/2} - t^{-n/2})(p^{n/2} + p^{-n/2}) \delta_{m+n,0}~, \qquad [\mb{P}, \mb{Q}] = 2\ ,
\end{align}
where $q, t,p \equiv qt^{-1}\in\mathbb{C}^\times$ are deformation parameters. The generators can be written as
\begin{align}
   \mb{T}(z) \equiv \sum_{m} \mb{T}_m z^{-m} = \mb{Y}(p^{-1/2}z) + \mb{Y}(p^{1/2}z)^{-1}, \qquad \mb{Y}(z) \equiv ~:\exp \Bigg[  \sum_{m\ne 0} \frac{\mb{a}_m z^{-m}}{p^{m/2} + p^{-m/2}}  \Bigg]:~,
\end{align}
where $:~~:$ denotes normal ordering (i.e. positive modes and $\mb{P}$ to the right). 

A special and important operator is the screening current $\mb{S}(x)$, having the property to commute with the $q$-Virasoro generators up to a total difference, namely $[\mb{T}_m,\mb{S}(x)]=\mb{O}_m(\lambda x)-\mb{O}(x)$ for all $m\in\mathbb{Z}$ and some operators $\mb{O}_m(x)$ and $\lambda\in\mathbb{C}$. A solution is
\begin{align}
   \mb{S}(x) \equiv \ : \exp \Bigg[ - \sum_{m \ne 0} \frac{\mb{a}_m x^{-m}}{q^{m/2} - q^{-m/2}} +\sqrt{\beta}\mb{Q}+(\sqrt{\beta} \mb{P}+1)\ln x\Bigg] : \ ,\qquad \beta\equiv \frac{\ln t}{\ln q}~,
\end{align}
fulfilling the screening property with $\lambda=q$.
Notice that the $q\leftrightarrow t^{-1}$ symmetry of the algebra implies the existence of another screening current $\widetilde{\mb{S}}(x)$ where the $q$ and $t^{-1}$ parameters are exchanged. In the following subsection, we review a construction which uses this symmetry as a starting point to introduce three mutually commuting $q$-Virasoro algebras admitting only three simultaneous and independent screening currents. 

\subsection{The modular triple}

\begin{figure}
\centering
\includegraphics[width=0.4\textwidth]{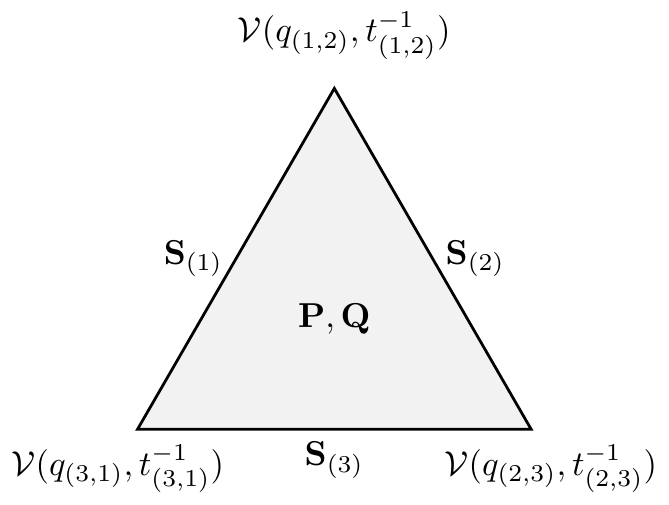}
\caption{\label{algebra-triangle}A pictorial representation of the gluing of three $q$-Virasoro algebras into the modular triple.}
\end{figure}

The modular triple construction for the $q$-Virasoro algebra ($\mathcal{V}(q,t^{-1})$ for short) starts with three copies $\mathcal{V}(q_{(i,j)}, t^{-1}_{(i,j)})$ with generators $\{\mb{T}_m^{(i,j)},m\in\mathbb{Z}\}$ and parameters $(q_{(i,j)},t_{(i,j)})$ taking values in Table \ref{parameter-table}, for $(i, j) \in \{(1,2), (2,3), (3,1)\}$. The corresponding Heisenberg algebras are denoted by $\{\mb{a}^{(i,j)}_m, \mb{P}^{(i,j)}, \mb{Q}^{(i,j)}\}$. To combine the three $q$-Virasoro algebras into the modular triple, we identify their zero modes $\mb{P}^{(i,j)}, \mb{Q}^{(i,j)}$ with the same operators $\mb{P}, \mb{Q}$ up to  proportionality constants. See Figure \ref{algebra-triangle} for an illustration. By direct computation, it is possible to check that there are three composite screening currents that simultaneously commute with all the generators up to total differences, which can compactly be presented as
\begin{multline}
  \mb{S}_{(i)}(X)\equiv \,  :\exp\Bigg[
    - \sum_{m \ne 0} \frac{
      (-1)^m \mb{a}^{(i,j)}_m \e^{-m \frac{2\pi \i X}{\omega_k}}}{\e^{\i\pi m \frac{\omega_j}{\omega_k}} - \e^{- \i\pi m \frac{\omega_j}{\omega_k}}} 
    - \sum_{m \ne 0} \frac{
      (-1)^m \mb{a}^{(k,i)}_m \e^{-m \frac{2\pi \i X}{\omega_j}}}{\e^{\i\pi m \frac{\omega_k}{\omega_j}} - \e^{- \i\pi m \frac{\omega_k}{\omega_j}}} +\\
    - \omega_i\mb{Q}+\frac{2 \pi \i (\mb{P}+\omega_{(j,k)})}{\omega_j\omega_k}X\Bigg]:\, ,
\end{multline}
where we defined $\omega_{(i,j)}\equiv \omega_i+\omega_j$. Note that each individual $\mb{S}_{(i)}(X)$ is a screening current of the modular double discussed in \cite{Nedelin:2016gwu}. 

\subsection{Modular triple and gauge theory partition functions}

One may guess that there is a hidden relation between the modular triple construction and the $S^5$ or $S^3_{(1)} \cup S^3_{(2)} \cup S^3_{(3)}$ partition functions by noting the identical equivariant parameters. To being with, let us consider the finite product of screening charges
\begin{align}\label{Zop}
  \mb{Z}_{n^{(1)},n^{(2)},n^{(3)}} \equiv \prod_{i=1,2,3}\frac{1}{n^{(i)}!}
    \Bigg[ \int\d X_{(i)}\,  \mb{S}_{(i)}(X_{(i)})\Bigg]^{n^{(i)}} \ ,
\end{align}
and introduce a Fock vacuum state $\ket{\xi}$ defined by $\mb{P}\ket{\xi} =\xi \ket{\xi}$, $\mb{a}^{(i,j)}_{m > 0} \ket{\xi}  = 0$. By construction, the action of the operator (\ref{Zop}) on the vacuum is formally annihilated by the generators of the modular triple, namely
\be\label{TZ}
\mb{T}^{(i,j)}_m \mb{Z}_{n^{(1)},n^{(2)},n^{(3)}}\ket{\xi}=0~,\quad m>0~.
\ee
Reorganizing the integrand of the operator (\ref{Zop}) using the commutation relations of the Heisenberg algebras and performing all the normal orderings, upon suitable identification of variables one can manifestly identify 
\be
\mb{Z}_{n^{(1)},n^{(2)},n^{(3)}} \ket{\xi} \simeq Z^{S^3_{(1)} \cup S^3_{(2)} \cup S^3_{(3)}}_{U(n^{(1)}), U(n^{(2)}), U(n^{(3)}), n_\text{f} = 0} (\tau^{(1,2)},\tau^{(2,3)},\tau^{(3,1)})~,
\ee
where the r.h.s. coincides with the matrix model of the gauge theory on $S^3_{(1)} \cup S^3_{(2)} \cup S^3_{(3)}$ discussed around (\ref{intmm}) in the absence of fundamental/anti-fundamental flavors, but with a potential parametrized by $\{\tau^{(i,j)}_n,n\in\mathbb{Z}_{>0}\}$ once the identifications $\mb{a}_{-n}^{(i,j)}\sim \tau_n^{(i,j)}$, $\mb{a}_{n}^{(i,j)}\sim \partial/\partial\tau_n^{(i,j)}$ for $n>0$ are made. This defines the generating function. The external momentum $\xi$ parametrizes the (non-independent) FI parameters
\be
-\i\sqrt{\omega_j\omega_k}\zeta^{(i)}_\text{FI}=\xi-\sum_{i=1,2,3}(n^{(i)}-1)\omega_i~.
\ee
The  inclusion of $N$ fundamental and anti-fundamental flavors can be obtained by shifting the times by additional parameters, namely
\be
\tau^{(i,j)}_{n} \to \tau^{(i,j)}_{n} +  \frac{1}{n}\sum_{A=1}^N  ( \e^{- \frac{2 \pi \i n}{\omega_k}(u_A+\omega/2)} - \e^{- \frac{2 \pi \i n}{\omega_k} (\tilde u_A +\omega/2)})~.
\ee
The additional factors recombine into the 1-loop determinants of fundamental and anti-fundamental chiral multiplets on the three $S^3$'s, whose masses are given by\footnote{It is easy to check that these masses satisfy precisely the same relations as in Footnote \ref{mass-relations}.}
\begin{align}
  m_s^{(i)} = \frac{\i u_s}{\sqrt{\omega_j \omega_k}} + \frac{\i}{2} \frac{\omega_i}{\sqrt{\omega_j \omega_k}}\ , \qquad \tilde m^{(i)}_s = \frac{\i u_s}{\sqrt{\omega_j \omega_k}} + \frac{\i}{2}\frac{\omega_i }{\sqrt{\omega_j \omega_k}} + \i (b_{(i)}+b_{(i)}^{-1})\ .
\end{align}

Finally, since the relation (\ref{TZ}) is valid for any set $\{n^{(i)}\}$ of non-negative integers, one can formally construct a much more general solution by considering an arbitrary superposition of matrix models (gran-canonical ensemble or coherent-like state), namely
\be
\ket{\mb{Z}}\equiv \sum_{\{n^{(i)}\geq 0\}}K^{n^{(1)},n^{(2)},n^{(3)}}\mb{Z}_{n^{(1)},n^{(2)},n^{(3)}}\ket{\xi}~,
\ee
for certain (oscillators/times-independent) coefficients $K^{n^{(1)},n^{(2)},n^{(3)}}$. In order to reproduce (\ref{finalresult}), one needs to set $K^{n^{(1)},n^{(2)},n^{(3)}}=1$, weight with the $S^5$ classical and free hyper 1-loop factors and sum over all the possible choices of HV (intermediate states). This construction yields a neat 2d CFT-like interpretation to the $S^5$ partition function, consistent with the BPS/CFT and 5d AGT expectations and open/closed duality of non-perturbative topological strings \cite{Lockhart:2012vp}.

\section{Discussion}\label{sec:disc}
Unfortunately, at present we do not have a satisfactory explanation of the exponentiation of the screening charges (i.e. $K^{n^{(1)},n^{(2)},n^{(3)}}=1$) nor of the weight factor without explicitly referring to some gauge theory argument. However, from the algebraic viewpoint, besides the construction of the proposal being very simple and natural, we may observe that: $i$) thanks to the exponentiation $\sum_{\{n^{(i)}\}} \mb{Z}_{n^{(1)}, n^{(2)},n^{(3)}}\sim \exp (\sum_{i}\int \mb{S}_{(i)})$, when this operator is sandwiched between external  Fock states with finite momentum (as necessary in order to have a pure defect partition function/Dotsenko-Fateev $q$-conformal block) the right number of screening charges is automatically picked up; also, the exponentiation hints to a worldsheet model similar to Liouville 2d CFT proposed in \cite{Nieri:2017vrb}; $ii$) the simple and specific weight factor may be needed once analytic, modular or crossing symmetry properties are imposed \cite{Nieri:2013yra}; $iii)$ the open questions may be solved by studying the more elementary Ding-Iohora-Miki algebra \cite{Ding1997,doi:10.1063/1.2823979}, the fundamental symmetry of matrix models arising in the present gauge theory context \cite{Mironov:2015thk,Awata:2016riz,Mironov:2016yue,Mironov:2016cyq}, which in our setup would  give rise to the $q$-Virasoro algebra together with and additional Heisenberg algebra \cite{2010arXiv1002.2485F} which may play an important role here (similarly to the undeformed 4d case \cite{Alba:2010qc}). What we have discussed in this short note deserve further studies, also in view of a possible parallel story on $S^4$ which we leave for future work.

\section*{Acknowledgments}
The research of M.Z. and F.N. is supported in part by Vetenskapsr\r{a}det under grant \#2014-5517, by the STINT grant and by the grant ``Geometry and Physics" from the Knut and Alice Wallenberg foundation. Y.P. is supported by the 100 Talents Program of Sun Yat-sen University under Grant No.74130-18831116.

  


\bibliographystyle{utphys}

\begin{thebibliography}{100}

\bibitem{Teschner:2016yzf}
J.~Teschner, ed., \href{http://dx.doi.org/10.1007/978-3-319-18769-3}{{\em {New
  Dualities of Supersymmetric Gauge Theories}}}.
\newblock Mathematical Physics Studies. Springer, Cham, Switzerland,
2016.
\newblock

\bibitem{Pestun:2016zxk}
V.~Pestun {\em et al.}, ``{Localization techniques in quantum field
  theories},'' \href{http://dx.doi.org/10.1088/1751-8121/aa63c1}{{\em J. Phys.}
  {\bf A50} (2017) no.~44, 440301},
\href{http://arxiv.org/abs/1608.02952}{{\tt arXiv:1608.02952 [hep-th]}}.

\bibitem{Pasquetti:2011fj}
S.~Pasquetti, ``{Factorisation of N = 2 Theories on the Squashed 3-Sphere},''
  \href{http://dx.doi.org/10.1007/JHEP04(2012)120}{{\em JHEP} {\bf 04} (2012)
  120},
\href{http://arxiv.org/abs/1111.6905}{{\tt arXiv:1111.6905 [hep-th]}}.

\bibitem{Doroud:2012xw}
N.~Doroud, J.~Gomis, B.~Le~Floch, and S.~Lee, ``{Exact Results in D=2
  Supersymmetric Gauge Theories},''
  \href{http://dx.doi.org/10.1007/JHEP05(2013)093}{{\em JHEP} {\bf 05} (2013)
  093},
\href{http://arxiv.org/abs/1206.2606}{{\tt arXiv:1206.2606 [hep-th]}}.

\bibitem{Benini:2012ui}
F.~Benini and S.~Cremonesi, ``{Partition Functions of ${\mathcal{N}=(2,2)}$
  Gauge Theories on S$^{2}$ and Vortices},''
  \href{http://dx.doi.org/10.1007/s00220-014-2112-z}{{\em Commun. Math. Phys.}
  {\bf 334} (2015) no.~3, 1483--1527},
\href{http://arxiv.org/abs/1206.2356}{{\tt arXiv:1206.2356 [hep-th]}}.

\bibitem{Fujitsuka:2013fga}
M.~Fujitsuka, M.~Honda, and Y.~Yoshida, ``{Higgs branch localization of 3d N =
  2 theories},'' \href{http://dx.doi.org/10.1093/ptep/ptu158}{{\em PTEP} {\bf
  2014} (2014) no.~12, 123B02},
\href{http://arxiv.org/abs/1312.3627}{{\tt arXiv:1312.3627 [hep-th]}}.

\bibitem{Benini:2013yva}
F.~Benini and W.~Peelaers, ``{Higgs branch localization in three dimensions},''
  \href{http://dx.doi.org/10.1007/JHEP05(2014)030}{{\em JHEP} {\bf 05} (2014)
  030},
\href{http://arxiv.org/abs/1312.6078}{{\tt arXiv:1312.6078 [hep-th]}}.

\bibitem{Taki:2013opa}
M.~Taki, ``{Holomorphic Blocks for 3d Non-abelian Partition Functions},''
\href{http://arxiv.org/abs/1303.5915}{{\tt arXiv:1303.5915 [hep-th]}}.

\bibitem{Peelaers:2014ima}
W.~Peelaers, ``{Higgs branch localization of $ \mathcal{N} $ = 1 theories on
  S$^{3}$ x S$^{1}$},'' \href{http://dx.doi.org/10.1007/JHEP08(2014)060}{{\em
  JHEP} {\bf 08} (2014)  060},
\href{http://arxiv.org/abs/1403.2711}{{\tt arXiv:1403.2711 [hep-th]}}.

\bibitem{Yoshida:2014qwa}
Y.~Yoshida, ``{Factorization of 4d N=1 superconformal index},''
\href{http://arxiv.org/abs/1403.0891}{{\tt arXiv:1403.0891 [hep-th]}}.

\bibitem{Pan:2014bwa}
Y.~Pan, ``{5d Higgs Branch Localization, Seiberg-Witten Equations and Contact
  Geometry},'' \href{http://dx.doi.org/10.1007/JHEP01(2015)145}{{\em JHEP} {\bf
  01} (2015)  145},
\href{http://arxiv.org/abs/1406.5236}{{\tt arXiv:1406.5236 [hep-th]}}.

\bibitem{Pan:2015hza}
Y.~Pan and W.~Peelaers, ``{Ellipsoid partition function from Seiberg-Witten
  monopoles},'' \href{http://dx.doi.org/10.1007/JHEP10(2015)183}{{\em JHEP}
  {\bf 10} (2015)  183},
\href{http://arxiv.org/abs/1508.07329}{{\tt arXiv:1508.07329 [hep-th]}}.

\bibitem{Chen:2015fta}
H.-Y. Chen and T.-H. Tsai, ``{On Higgs branch localization of Seiberg--Witten
  theories on an ellipsoid},''
  \href{http://dx.doi.org/10.1093/ptep/ptv188}{{\em PTEP} {\bf 2016} (2016)
  no.~1, 013B09},
\href{http://arxiv.org/abs/1506.04390}{{\tt arXiv:1506.04390 [hep-th]}}.

\bibitem{Pestun:2007rz}
V.~Pestun, ``{Localization of gauge theory on a four-sphere and supersymmetric
  Wilson loops},'' \href{http://dx.doi.org/10.1007/s00220-012-1485-0}{{\em
  Commun. Math. Phys.} {\bf 313} (2012)  71--129},
\href{http://arxiv.org/abs/0712.2824}{{\tt arXiv:0712.2824 [hep-th]}}.

\bibitem{Hama:2012bg}
N.~Hama and K.~Hosomichi, ``{Seiberg-Witten Theories on Ellipsoids},''
  \href{http://dx.doi.org/10.1007/JHEP09(2012)033,
  10.1007/JHEP10(2012)051}{{\em JHEP} {\bf 09} (2012)  033},
  \href{http://arxiv.org/abs/1206.6359}{{\tt arXiv:1206.6359 [hep-th]}}.
[Addendum: JHEP10,051(2012)].

\bibitem{Alday:2009aq}
L.~F. Alday, D.~Gaiotto, and Y.~Tachikawa, ``{Liouville Correlation Functions
  from Four-dimensional Gauge Theories},''
  \href{http://dx.doi.org/10.1007/s11005-010-0369-5}{{\em Lett. Math. Phys.}
  {\bf 91} (2010)  167--197},
\href{http://arxiv.org/abs/0906.3219}{{\tt arXiv:0906.3219 [hep-th]}}.

\bibitem{Wyllard:2009hg}
N.~Wyllard, ``{A(N-1) conformal Toda field theory correlation functions from
  conformal N = 2 SU(N) quiver gauge theories},''
  \href{http://dx.doi.org/10.1088/1126-6708/2009/11/002}{{\em JHEP} {\bf 11}
  (2009)  002},
\href{http://arxiv.org/abs/0907.2189}{{\tt arXiv:0907.2189 [hep-th]}}.

\bibitem{Gomis:2016ljm}
J.~Gomis, B.~Le~Floch, Y.~Pan, and W.~Peelaers, ``{Intersecting Surface Defects
  and Two-Dimensional CFT},''
\href{http://arxiv.org/abs/1610.03501}{{\tt arXiv:1610.03501 [hep-th]}}.

\bibitem{Kallen:2012va}
J.~K{\"a}ll{\'e}n, J.~Qiu, and M.~Zabzine, ``{The perturbative partition
  function of supersymmetric 5D Yang-Mills theory with matter on the
  five-sphere},'' \href{http://dx.doi.org/10.1007/JHEP08(2012)157}{{\em JHEP}
  {\bf 08} (2012)  157},
\href{http://arxiv.org/abs/1206.6008}{{\tt arXiv:1206.6008 [hep-th]}}.

\bibitem{Kallen:2012cs}
J.~K{\"a}ll{\'e}n and M.~Zabzine, ``{Twisted supersymmetric 5D Yang-Mills
  theory and contact geometry},''
  \href{http://dx.doi.org/10.1007/JHEP05(2012)125}{{\em JHEP} {\bf 05} (2012)
  125},
\href{http://arxiv.org/abs/1202.1956}{{\tt arXiv:1202.1956 [hep-th]}}.

\bibitem{Kim:2012qf}
H.-C. Kim, J.~Kim, and S.~Kim, ``{Instantons on the 5-sphere and M5-branes},''
\href{http://arxiv.org/abs/1211.0144}{{\tt arXiv:1211.0144 [hep-th]}}.

\bibitem{Hosomichi:2012ek}
K.~Hosomichi, R.-K. Seong, and S.~Terashima, ``{Supersymmetric Gauge Theories
  on the Five-Sphere},''
  \href{http://dx.doi.org/10.1016/j.nuclphysb.2012.08.007}{{\em Nucl. Phys.}
  {\bf B865} (2012)  376--396},
\href{http://arxiv.org/abs/1203.0371}{{\tt arXiv:1203.0371 [hep-th]}}.

\bibitem{Imamura:2012bm}
Y.~Imamura, ``{Perturbative partition function for squashed $S^5$},''
\href{http://arxiv.org/abs/1210.6308}{{\tt arXiv:1210.6308 [hep-th]}}.

\bibitem{Lockhart:2012vp}
G.~Lockhart and C.~Vafa, ``{Superconformal Partition Functions and
  Non-perturbative Topological Strings},''
\href{http://arxiv.org/abs/1210.5909}{{\tt arXiv:1210.5909 [hep-th]}}.

\bibitem{Pan:2016fbl}
Y.~Pan and W.~Peelaers, ``{Intersecting Surface Defects and Instanton Partition
  Functions},''
\href{http://arxiv.org/abs/1612.04839}{{\tt arXiv:1612.04839 [hep-th]}}.

\bibitem{Aharony:1997bh}
O.~Aharony, A.~Hanany, and B.~Kol, ``{Webs of (p,q) five-branes,
  five-dimensional field theories and grid diagrams},''
  \href{http://dx.doi.org/10.1088/1126-6708/1998/01/002}{{\em JHEP} {\bf 01}
  (1998)  002},
\href{http://arxiv.org/abs/hep-th/9710116}{{\tt arXiv:hep-th/9710116
  [hep-th]}}.

\bibitem{Aharony:1997ju}
O.~Aharony and A.~Hanany, ``{Branes, superpotentials and superconformal fixed
  points},'' \href{http://dx.doi.org/10.1016/S0550-3213(97)00472-0}{{\em Nucl.
  Phys.} {\bf B504} (1997)  239--271},
\href{http://arxiv.org/abs/hep-th/9704170}{{\tt arXiv:hep-th/9704170
  [hep-th]}}.

\bibitem{Awata:2009ur}
H.~Awata and Y.~Yamada, ``{Five-dimensional AGT Conjecture and the Deformed
  Virasoro Algebra},'' \href{http://dx.doi.org/10.1007/JHEP01(2010)125}{{\em
  JHEP} {\bf 01} (2010)  125},
\href{http://arxiv.org/abs/0910.4431}{{\tt arXiv:0910.4431 [hep-th]}}.

\bibitem{Awata:2010yy}
H.~Awata and Y.~Yamada, ``{Five-dimensional AGT Relation and the Deformed
  beta-ensemble},'' \href{http://dx.doi.org/10.1143/PTP.124.227}{{\em Prog.
  Theor. Phys.} {\bf 124} (2010)  227--262},
\href{http://arxiv.org/abs/1004.5122}{{\tt arXiv:1004.5122 [hep-th]}}.

\bibitem{Awata:2011dc}
H.~Awata, B.~Feigin, A.~Hoshino, M.~Kanai, J.~Shiraishi, and S.~Yanagida,
  ``{Notes on Ding-Iohara algebra and AGT conjecture},''
\href{http://arxiv.org/abs/1106.4088}{{\tt arXiv:1106.4088 [math-ph]}}.

\bibitem{Mironov:2011dk}
A.~Mironov, A.~Morozov, S.~Shakirov, and A.~Smirnov, ``{Proving AGT conjecture
  as HS duality: extension to five dimensions},''
  \href{http://dx.doi.org/10.1016/j.nuclphysb.2011.09.021}{{\em Nucl. Phys.}
  {\bf B855} (2012)  128--151},
\href{http://arxiv.org/abs/1105.0948}{{\tt arXiv:1105.0948 [hep-th]}}.

\bibitem{Nekrasov:2012xe}
N.~Nekrasov and V.~Pestun, ``{Seiberg-Witten geometry of four dimensional N=2
  quiver gauge theories},''
\href{http://arxiv.org/abs/1211.2240}{{\tt arXiv:1211.2240 [hep-th]}}.

\bibitem{Nekrasov:2013xda}
N.~Nekrasov, V.~Pestun, and S.~Shatashvili, ``{Quantum geometry and quiver
  gauge theories},''
\href{http://arxiv.org/abs/1312.6689}{{\tt arXiv:1312.6689 [hep-th]}}.

\bibitem{Carlsson:2013jka}
E.~Carlsson, N.~Nekrasov, and A.~Okounkov, ``{Five dimensional gauge theories
  and vertex operators},'' {\em Moscow Math. J.} {\bf 14} (2014) no.~1, 39--61,
\href{http://arxiv.org/abs/1308.2465}{{\tt arXiv:1308.2465 [math.RT]}}.

\bibitem{Nieri:2013yra}
F.~Nieri, S.~Pasquetti, and F.~Passerini, ``{3d and 5d Gauge Theory Partition
  Functions as $q$-deformed CFT Correlators},''
  \href{http://dx.doi.org/10.1007/s11005-014-0727-9}{{\em Lett. Math. Phys.}
  {\bf 105} (2015) no.~1, 109--148},
\href{http://arxiv.org/abs/1303.2626}{{\tt arXiv:1303.2626 [hep-th]}}.

\bibitem{Nieri:2013vba}
F.~Nieri, S.~Pasquetti, F.~Passerini, and A.~Torrielli, ``{5D partition
  functions, q-Virasoro systems and integrable spin-chains},''
  \href{http://dx.doi.org/10.1007/JHEP12(2014)040}{{\em JHEP} {\bf 12} (2014)
  040},
\href{http://arxiv.org/abs/1312.1294}{{\tt arXiv:1312.1294 [hep-th]}}.

\bibitem{Aganagic:2013tta}
M.~Aganagic, N.~Haouzi, C.~Kozcaz, and S.~Shakirov, ``{Gauge/Liouville
  Triality},''
\href{http://arxiv.org/abs/1309.1687}{{\tt arXiv:1309.1687 [hep-th]}}.

\bibitem{Aganagic:2014kja}
M.~Aganagic and S.~Shakirov, ``{Gauge/Vortex duality and AGT},''
\href{http://arxiv.org/abs/1412.7132}{{\tt arXiv:1412.7132 [hep-th]}}.

\bibitem{Aganagic:2014oia}
M.~Aganagic, N.~Haouzi, and S.~Shakirov, ``{$A_n$-Triality},''
\href{http://arxiv.org/abs/1403.3657}{{\tt arXiv:1403.3657 [hep-th]}}.

\bibitem{Zenkevich:2014lca}
Y.~Zenkevich, ``Generalized macdonald polynomials, spectral duality for
  conformal blocks and agt correspondence in five dimensions,''
  \href{http://dx.doi.org/10.1007/JHEP05(2015)131}{{\em JHEP} {\bf 05} (2015)
  131},
\href{http://arxiv.org/abs/1412.8592}{{\tt arXiv:1412.8592 [hep-th]}}.

\bibitem{Mitev:2014isa}
V.~Mitev and E.~Pomoni, ``{Toda 3-Point Functions From Topological Strings},''
  \href{http://dx.doi.org/10.1007/JHEP06(2015)049}{{\em JHEP} {\bf 06} (2015)
  049},
\href{http://arxiv.org/abs/1409.6313}{{\tt arXiv:1409.6313 [hep-th]}}.

\bibitem{Isachenkov:2014eya}
M.~Isachenkov, V.~Mitev, and E.~Pomoni, ``{Toda 3-Point Functions From
  Topological Strings II},''
  \href{http://dx.doi.org/10.1007/JHEP08(2016)066}{{\em JHEP} {\bf 08} (2016)
  066},
\href{http://arxiv.org/abs/1412.3395}{{\tt arXiv:1412.3395 [hep-th]}}.

\bibitem{Aganagic:2015cta}
M.~Aganagic and N.~Haouzi, ``{ADE Little String Theory on a Riemann Surface
  (and Triality)},''
\href{http://arxiv.org/abs/1506.04183}{{\tt arXiv:1506.04183 [hep-th]}}.

\bibitem{Kimura:2015rgi}
T.~Kimura and V.~Pestun, ``{Quiver W-algebras},''
  \href{http://dx.doi.org/10.1007/s11005-018-1072-1}{{\em Lett. Math. Phys.}
  {\bf 108} (2018) no.~6, 1351--1381},
\href{http://arxiv.org/abs/1512.08533}{{\tt arXiv:1512.08533 [hep-th]}}.

\bibitem{Benvenuti:2016dcs}
S.~Benvenuti, G.~Bonelli, M.~Ronzani, and A.~Tanzini, ``{Symmetry enhancements
  via 5d instantons, $ q\mathcal{W} $ -algebrae and (1, 0) superconformal
  index},'' \href{http://dx.doi.org/10.1007/JHEP09(2016)053}{{\em JHEP} {\bf
  09} (2016)  053},
\href{http://arxiv.org/abs/1606.03036}{{\tt arXiv:1606.03036 [hep-th]}}.

\bibitem{Bourgine:2016vsq}
J.-E. Bourgine, M.~Fukuda, Y.~Matsuo, H.~Zhang, and R.-D. Zhu, ``{Coherent
  states in quantum $\mathcal{W}_{1+\infty}$ algebra and qq-character for 5d
  Super Yang-Mills},'' \href{http://dx.doi.org/10.1093/ptep/ptw165}{{\em PTEP}
  {\bf 2016} (2016) no.~12, 123B05},
\href{http://arxiv.org/abs/1606.08020}{{\tt arXiv:1606.08020 [hep-th]}}.

\bibitem{Bourgine:2017jsi}
J.-E. Bourgine, M.~Fukuda, K.~Harada, Y.~Matsuo, and R.-D. Zhu, ``{(p,q)-webs
  of DIM representations, 5d N=1 instanton partition functions and
  qq-characters},'' \href{http://dx.doi.org/10.1007/JHEP11(2017)034}{{\em JHEP}
  {\bf 11} (2017)  034},
\href{http://arxiv.org/abs/1703.10759}{{\tt arXiv:1703.10759 [hep-th]}}.

\bibitem{Aganagic:2017smx}
M.~Aganagic, E.~Frenkel, and A.~Okounkov, ``{Quantum q-Langlands
  Correspondence},''
\href{http://arxiv.org/abs/1701.03146}{{\tt arXiv:1701.03146 [hep-th]}}.

\bibitem{Katz:1997eq}
S.~Katz, P.~Mayr, and C.~Vafa, ``{Mirror symmetry and exact solution of 4D N=2
  gauge theories: 1.},''
  \href{http://dx.doi.org/10.4310/ATMP.1997.v1.n1.a2}{{\em Adv. Theor. Math.
  Phys.} {\bf 1} (1998)  53--114},
\href{http://arxiv.org/abs/hep-th/9706110}{{\tt arXiv:hep-th/9706110
  [hep-th]}}.

\bibitem{Bao:2011rc}
L.~Bao, E.~Pomoni, M.~Taki, and F.~Yagi, ``{M5-Branes, Toric Diagrams and Gauge
  Theory Duality},'' \href{http://dx.doi.org/10.1007/JHEP04(2012)105}{{\em
  JHEP} {\bf 04} (2012)  105},
\href{http://arxiv.org/abs/1112.5228}{{\tt arXiv:1112.5228 [hep-th]}}.

\bibitem{Gaiotto:2012xa}
D.~Gaiotto, L.~Rastelli, and S.~S. Razamat, ``{Bootstrapping the superconformal
  index with surface defects},''
  \href{http://dx.doi.org/10.1007/JHEP01(2013)022}{{\em JHEP} {\bf 01} (2013)
  022},
\href{http://arxiv.org/abs/1207.3577}{{\tt arXiv:1207.3577 [hep-th]}}.

\bibitem{Chen:2014rca}
H.-Y. Chen and H.-Y. Chen, ``{Heterotic Surface Defects and Dualities from
  2d/4d Indices},'' \href{http://dx.doi.org/10.1007/JHEP10(2014)004}{{\em JHEP}
  {\bf 10} (2014)  004},
\href{http://arxiv.org/abs/1407.4587}{{\tt arXiv:1407.4587 [hep-th]}}.

\bibitem{Alday:2009fs}
L.~F. Alday, D.~Gaiotto, S.~Gukov, Y.~Tachikawa, and H.~Verlinde, ``{Loop and
  surface operators in N=2 gauge theory and Liouville modular geometry},''
  \href{http://dx.doi.org/10.1007/JHEP01(2010)113}{{\em JHEP} {\bf 01} (2010)
  113},
\href{http://arxiv.org/abs/0909.0945}{{\tt arXiv:0909.0945 [hep-th]}}.

\bibitem{Bonelli:2011wx}
G.~Bonelli, A.~Tanzini, and J.~Zhao, ``{The Liouville side of the Vortex},''
  \href{http://dx.doi.org/10.1007/JHEP09(2011)096}{{\em JHEP} {\bf 09} (2011)
  096},
\href{http://arxiv.org/abs/1107.2787}{{\tt arXiv:1107.2787 [hep-th]}}.

\bibitem{Bonelli:2011fq}
G.~Bonelli, A.~Tanzini, and J.~Zhao, ``{Vertices, Vortices and Interacting
  Surface Operators},'' \href{http://dx.doi.org/10.1007/JHEP06(2012)178}{{\em
  JHEP} {\bf 06} (2012)  178},
\href{http://arxiv.org/abs/1102.0184}{{\tt arXiv:1102.0184 [hep-th]}}.

\bibitem{Gomis:2014eya}
J.~Gomis and B.~Le~Floch, ``{M2-brane surface operators and gauge theory
  dualities in Toda},'' \href{http://dx.doi.org/10.1007/JHEP04(2016)183}{{\em
  JHEP} {\bf 04} (2016)  183},
\href{http://arxiv.org/abs/1407.1852}{{\tt arXiv:1407.1852 [hep-th]}}.

\bibitem{Bullimore:2016hdc}
M.~Bullimore, T.~Dimofte, D.~Gaiotto, J.~Hilburn, and H.-C. Kim, ``{Vortices
  and Vermas},''
\href{http://arxiv.org/abs/1609.04406}{{\tt arXiv:1609.04406 [hep-th]}}.

\bibitem{Nieri:2017ntx}
F.~Nieri, Y.~Pan, and M.~Zabzine, ``{3d Expansions of 5d Instanton Partition
  Functions},'' \href{http://dx.doi.org/10.1007/JHEP04(2018)092}{{\em JHEP}
  {\bf 04} (2018)  092},
\href{http://arxiv.org/abs/1711.06150}{{\tt arXiv:1711.06150 [hep-th]}}.

\bibitem{Nieri:2017vrb}
F.~Nieri, Y.~Pan, and M.~Zabzine, ``{$q$-Virasoro modular triple},''
\href{http://arxiv.org/abs/1710.07170}{{\tt arXiv:1710.07170 [hep-th]}}.

\bibitem{Coman:2017qgv}
I.~Coman, E.~Pomoni, and J.~Teschner, ``{Toda conformal blocks, quantum groups,
  and flat connections},''
\href{http://arxiv.org/abs/1712.10225}{{\tt arXiv:1712.10225 [hep-th]}}.

\bibitem{Qiu:2014oqa}
J.~Qiu, L.~Tizzano, J.~Winding, and M.~Zabzine, ``{Gluing Nekrasov partition
  functions},'' \href{http://dx.doi.org/10.1007/s00220-015-2351-7}{{\em Commun.
  Math. Phys.} {\bf 337} (2015) no.~2, 785--816},
\href{http://arxiv.org/abs/1403.2945}{{\tt arXiv:1403.2945 [hep-th]}}.

\bibitem{Hama:2011ea}
N.~Hama, K.~Hosomichi, and S.~Lee, ``{SUSY Gauge Theories on Squashed
  Three-Spheres},'' \href{http://dx.doi.org/10.1007/JHEP05(2011)014}{{\em JHEP}
  {\bf 05} (2011)  014},
\href{http://arxiv.org/abs/1102.4716}{{\tt arXiv:1102.4716 [hep-th]}}.

\bibitem{Narukawa:2003}
A.~Narukawa, ``{The modular properties and the integral representations of the
  multiple elliptic gamma functions},''
\href{http://arxiv.org/abs/math/0306164}{{\tt arXiv:math/0306164 [math.QA]}}.

\bibitem{Qiu:2013aga}
J.~Qiu and M.~Zabzine, ``{Factorization of 5D super Yang-Mills theory on
  $Y^{p,q}$ spaces},'' \href{http://dx.doi.org/10.1103/PhysRevD.89.065040}{{\em
  Phys. Rev.} {\bf D89} (2014) no.~6, 065040},
\href{http://arxiv.org/abs/1312.3475}{{\tt arXiv:1312.3475 [hep-th]}}.

\bibitem{Lossev:1997xxx}
A.~Lossev, N.~Nekrasov, and S.~L. Shatashvili, ``{Testing Seiberg-Witten
  solution},''
\href{http://arxiv.org/abs/9801061}{{\tt arXiv:9801061 [hep-th]}}.

\bibitem{Moore:1998et}
G.~W. Moore, N.~Nekrasov, and S.~Shatashvili, ``{D particle bound states and
  generalized instantons},''
  \href{http://dx.doi.org/10.1007/s002200050016}{{\em Commun. Math. Phys.} {\bf
  209} (2000)  77--95},
\href{http://arxiv.org/abs/hep-th/9803265}{{\tt arXiv:hep-th/9803265
  [hep-th]}}.

\bibitem{Moore:1997dj}
G.~W. Moore, N.~Nekrasov, and S.~Shatashvili, ``{Integrating over Higgs
  branches},'' \href{http://dx.doi.org/10.1007/PL00005525}{{\em Commun. Math.
  Phys.} {\bf 209} (2000)  97--121},
\href{http://arxiv.org/abs/hep-th/9712241}{{\tt arXiv:hep-th/9712241
  [hep-th]}}.

\bibitem{Losev:1997tp}
A.~Losev, N.~Nekrasov, and S.~L. Shatashvili, ``{Issues in topological gauge
  theory},'' \href{http://dx.doi.org/10.1016/S0550-3213(98)00628-2}{{\em Nucl.
  Phys.} {\bf B534} (1998)  549--611},
\href{http://arxiv.org/abs/hep-th/9711108}{{\tt arXiv:hep-th/9711108
  [hep-th]}}.

\bibitem{Nekrasov:2002qd}
N.~A. Nekrasov, ``{Seiberg-Witten prepotential from instanton counting},''
  \href{http://dx.doi.org/10.4310/ATMP.2003.v7.n5.a4}{{\em Adv. Theor. Math.
  Phys.} {\bf 7} (2004)  831--864},
\href{http://arxiv.org/abs/hep-th/0206161}{{\tt arXiv:hep-th/0206161
  [hep-th]}}.

\bibitem{Nekrasov:2003rj}
N.~Nekrasov and A.~Okounkov, ``{Seiberg-Witten theory and random partitions},''
\href{http://arxiv.org/abs/hep-th/0306238}{{\tt arXiv:hep-th/0306238
  [hep-th]}}.

\bibitem{Gaiotto:2014ina}
D.~Gaiotto and H.-C. Kim, ``{Surface defects and instanton partition
  functions},'' \href{http://dx.doi.org/10.1007/JHEP10(2016)012}{{\em JHEP}
  {\bf 10} (2016)  012},
\href{http://arxiv.org/abs/1412.2781}{{\tt arXiv:1412.2781 [hep-th]}}.

\bibitem{Nieri:2018xxx}
F.~Nieri, Y.~Pan, and M.~Zabzine, ``{To appear},''.

\bibitem{Kimura:2016ebq}
T.~Kimura, ``{Double quantization of Seiberg-Witten geometry and W-algebras},''
\href{http://arxiv.org/abs/1612.07590}{{\tt arXiv:1612.07590 [hep-th]}}.

\bibitem{Morozov:1994hh}
A.~Morozov, ``{Integrability and matrix models},''
  \href{http://dx.doi.org/10.1070/PU1994v037n01ABEH000001}{{\em Phys. Usp.}
  {\bf 37} (1994)  1--55},
\href{http://arxiv.org/abs/hep-th/9303139}{{\tt arXiv:hep-th/9303139
  [hep-th]}}.

\bibitem{Nekrasov:2015wsu}
N.~Nekrasov, ``{BPS/CFT correspondence: non-perturbative Dyson-Schwinger
  equations and qq-characters},''
  \href{http://dx.doi.org/10.1007/JHEP03(2016)181}{{\em JHEP} {\bf 03} (2016)
  181},
\href{http://arxiv.org/abs/1512.05388}{{\tt arXiv:1512.05388 [hep-th]}}.

\bibitem{Kim:2016qqs}
H.-C. Kim, ``{Line defects and 5d instanton partition functions},''
  \href{http://dx.doi.org/10.1007/JHEP03(2016)199}{{\em JHEP} {\bf 03} (2016)
  199},
\href{http://arxiv.org/abs/1601.06841}{{\tt arXiv:1601.06841 [hep-th]}}.

\bibitem{Assel:2018rcw}
B.~Assel and A.~Sciarappa, ``{Wilson loops in 5d $\mathcal{N}=1$ theories and
  S-duality},''
\href{http://arxiv.org/abs/1806.09636}{{\tt arXiv:1806.09636 [hep-th]}}.

\bibitem{Nedelin:2016gwu}
A.~Nedelin, F.~Nieri, and M.~Zabzine, ``{$q$-Virasoro modular double and 3d
  partition functions},''
  \href{http://dx.doi.org/10.1007/s00220-017-2882-1}{{\em Commun. Math. Phys.}
  {\bf 353} (2017) no.~3, 1059--1102},
\href{http://arxiv.org/abs/1605.07029}{{\tt arXiv:1605.07029 [hep-th]}}.

\bibitem{Shiraishi:1995rp}
J.~Shiraishi, H.~Kubo, H.~Awata, and S.~Odake, ``{A Quantum deformation of the
  Virasoro algebra and the Macdonald symmetric functions},''
  \href{http://dx.doi.org/10.1007/BF00398297}{{\em Lett. Math. Phys.} {\bf 38}
  (1996)  33--51},
\href{http://arxiv.org/abs/q-alg/9507034}{{\tt arXiv:q-alg/9507034 [q-alg]}}.

\bibitem{1997q.alg.....8006F}
E.~Frenkel and N.~Reshetikhin, ``{Deformations of W-algebras associated to
  simple Lie algebras},''
\href{http://arxiv.org/abs/9708006}{{\tt arXiv:9708006 [q-alg]}}.

\bibitem{Ding1997}
J.~Ding and K.~Iohara,
  \href{http://dx.doi.org/10.1023/A:1007341410987}{``Generalization of drinfeld
  quantum affine algebras,''{\em Letters in Mathematical Physics} {\bf 41}
  (Jul, 1997)  181--193}. \url{https://doi.org/10.1023/A:1007341410987}.

\bibitem{doi:10.1063/1.2823979}
K.~Miki, ``A $(q,\gamma)$ analog of the $w_{1+\infty}$ˆž algebra,''
  \href{http://dx.doi.org/10.1063/1.2823979}{{\em Journal of Mathematical
  Physics} {\bf 48} (2007) no.~12, 123520},
  \href{http://arxiv.org/abs/https://doi.org/10.1063/1.2823979}{{\tt
  https://doi.org/10.1063/1.2823979}}. \url{https://doi.org/10.1063/1.2823979}.

\bibitem{Mironov:2015thk}
A.~Mironov, A.~Morozov, and Y.~Zenkevich, ``{On elementary proof of AGT
  relations from six dimensions},''
  \href{http://dx.doi.org/10.1016/j.physletb.2016.03.006}{{\em Phys. Lett.}
  {\bf B756} (2016)  208--211},
\href{http://arxiv.org/abs/1512.06701}{{\tt arXiv:1512.06701 [hep-th]}}.

\bibitem{Awata:2016riz}
H.~Awata, H.~Kanno, T.~Matsumoto, A.~Mironov, A.~Morozov, A.~Morozov,
  Y.~Ohkubo, and Y.~Zenkevich, ``{Explicit examples of DIM constraints for
  network matrix models},''
\href{http://arxiv.org/abs/1604.08366}{{\tt arXiv:1604.08366 [hep-th]}}.

\bibitem{Mironov:2016yue}
A.~Mironov, A.~Morozov, and Y.~Zenkevich, ``{Ding-Iohara-Miki symmetry of
  network matrix models},''
\href{http://arxiv.org/abs/1603.05467}{{\tt arXiv:1603.05467 [hep-th]}}.

\bibitem{Mironov:2016cyq}
A.~Mironov, A.~Morozov, and Y.~Zenkevich, ``Spectral duality in elliptic
  systems, six-dimensional gauge theories and topological strings,''
\href{http://arxiv.org/abs/1603.00304}{{\tt arXiv:1603.00304 [hep-th]}}.

\bibitem{2010arXiv1002.2485F}
B.~{Feigin}, A.~{Hoshino}, J.~{Shibahara}, J.~{Shiraishi}, and S.~{Yanagida},
  ``{Kernel function and quantum algebras},''
  \href{http://arxiv.org/abs/1002.2485}{{\tt arXiv:1002.2485 [math.QA]}}.

\bibitem{Alba:2010qc}
V.~A. Alba, V.~A. Fateev, A.~V. Litvinov, and G.~M. Tarnopolskiy, ``{On
  combinatorial expansion of the conformal blocks arising from AGT
  conjecture},'' \href{http://dx.doi.org/10.1007/s11005-011-0503-z}{{\em Lett.
  Math. Phys.} {\bf 98} (2011)  33--64},
\href{http://arxiv.org/abs/1012.1312}{{\tt arXiv:1012.1312 [hep-th]}}.

\end{thebibliography}

\providecommand{\href}[2]{#2}\begingroup\raggedright\endgroup

\end{document}